# Large topological Hall effect arising from spin reorientation in kagome magnet Fe$_3$Ge

Zixuan Zhang, Mingyue Zhao, Li Ma*, Guoke Li, Congmian Zhen, Dewei Zhao, and Denglu Hou

*Hebei Key Laboratory of Photophysics Research and Application, College of Physics, Hebei Normal University, Shijiazhuang 050024, China*

**Abstract**

Materials systems with spin chirality can provide ultra-high-density, ultra-fast, and ultra-low-power information carriers for digital transformation. These material systems include magnetic skyrmions, chiral domain walls, spin reorientation, and so on. The topological Hall effect (THE) has been identified as the most convenient and effective tool for detecting the presence of spin chirality in these systems. The research on the THE that may arise from spin reorientation and specifically in Fe$_3$Ge with spin reorientation remains an unexplored area, so we study the THE in Fe$_3$Ge Conduct systematic research. X-Ray Diffraction (XRD) results indicate that our Fe$_3$Ge ribbon sample has a D0$_{19}$ structure. First-principles calculations and magnetic and electrical testing confirm spin reorientation in the Fe$_3$Ge ribbon sample at 350 K. The Hall resistivity test results are consistent with our expectations, indicating the presence of the THE in the Fe$_3$Ge ribbon sample. The topological Hall resistivity reaches a maximum value of 0.69 μΩ cm at 400 K. For the first time, a detailed experimental study of the THE in Fe$_3$Ge with spin reorientation has been conducted, introducing a new member to the family of THE.

**Introduction**

The topological Hall effect (THE) is a transport-related topological property of magnetic materials and is a new member of the Hall effect family. Different from the anomalous Hall effect (AHE), THE is related to scalar spin chirality, breaks the time reversal symmetry and introduces Berry curvature of real space, without the need for spin-orbit coupling[1]. To realize the topological Hall effect, it is necessary to consider the spin-related Hamiltonian, which is:

$$H_s = \sum_{ij} J_{ij} S_i \cdot S_j + \sum_{ij} D_{ij} \cdot (S_i \times S_j) + + \sum_i K_i (e_i \cdot S_i)^2$$

the first term is Heisenberg exchange interaction, the second term is Dzyaloshinskii-Moriya interaction (DMI), and the third term is magnetocrystalline anisotropy. Controlling the topological spin structure mainly starts from the above items.

THE was originally discovered in non-centrosymmetric B20 systems such as MnSi[2], Mn$_{1-x}$Fe$_x$Si[3], and MnGe[4]. Large DMI lead to the emergence of non-coplanar spin structures in noncentrosymmetric systems. When this non-coplanar spin structure is coupled with the topological band structure of the material, a topological Hall phase can be formed. Therefore, some researchers pointed out that the bulk asymmetry-DMI is a mechanism to realize the topological Hall effect. And THE driven by interface DMI is realized in the multi-layer film structure of Ir/Fe/Co/Pt[5] and the double-layer structure of SrRuO$_3$ (SRO)-SrIrO$_3$ (SIO)[6]. In addition, some researchers have found that in $Mn_2RhSn$[7] and frustrated systems such as $Nd_2Mo_2O_7$[8], $Pr_2Ir_2O_7$[9], etc., they only need to control the Heisenberg exchange interaction to deflect the spin direction and form a non-coplanar structure. Then THE is realized, which greatly expands the means of realizing THE. Recently, researchers have recently observed THE in $NdCo_{5-x}Cu_x$[10] and $Fe_3Sn_2$[11] systems with spin reorientation. The spin reorientation produces non-coplanar spin configurations, which provides a new perspective on the origin of THE. However, there are still some limitations in the research on THE in spin reorientation systems, including the material system is not rich enough, etc. Through a survey of the literature, it was found that $Fe_3Ge$ has a spin reorientation temperature of 380 K. In addition, it has a Curie temperature (640 K) that is much higher than room temperature, which is very beneficial for applications.

In this work, $Fe_3Ge$ was used as a research platform to systematically study the relationship between spin reorientation and THE. A series of magnetic and electrical test results confirmed that the spin reorientation temperature of $Fe_3Ge$ sample is 350 K. The topological Hall resistivity increases rapidly after 350 K and reaches a maximum value of 0.69 μΩ cm at 400 K.

**Experimental method and theorical calculation details**

$Fe_3Ge$ polycrystalline ingots are prepared from Fe and Ge elements with a purity of more than 99.9 % using arc melting technology in an argon environment. According to the phase diagram[12], $Fe_3Ge$ has an cubic structure below 700 °C and a hexagonal structure at high temperatures. In order to obtain a pure hexagonal structure, we adopted the method of rapid solidification by ribbon spinnin. The $Fe_3Ge$ ribbon sample was obtained by placing the smelted block into a quartz tube with a small hole at the bottom, installing it in a ribbon spinning machine, and spraying the induction melted sample onto a high-speed rotating copper wheel in an argon environment, the surface speed of the copper wheel is 12.5 m/s. The structure of the $Fe_3Ge$ sample was characterized by X-ray diffractometer. The phase structure and lattice constants of the samples were obtained by the Rietveld fitting method in HighScore Plus software (PANalytical B.V.). DC testing of the samples was performed by a magnetic measurement system (MPMS, Quantum Design). AC susceptibility, longitudinal resistivity, and Hall resistivity were measured by a Physical Properties Test System (PPMS, Quantum Design).

Density functional theory calculations were performed using the generalized gradient approximation (GGA) and projector augmented wave (PAW) methods in the Vienna Ab initio Simulation Package (VASP). For the Brillouin zone integration, a G-centered k-point grid of 11×11×12 was used with a plane-wave cutoff energy of 450 eV. Based on the known crystal structure, the PBE (Perdew-Burke-Ernzerhof) potential and spin-orbit coupling (SOC) were used to perform energy comparisons between different spin configurations.

# Experimental results and discussion

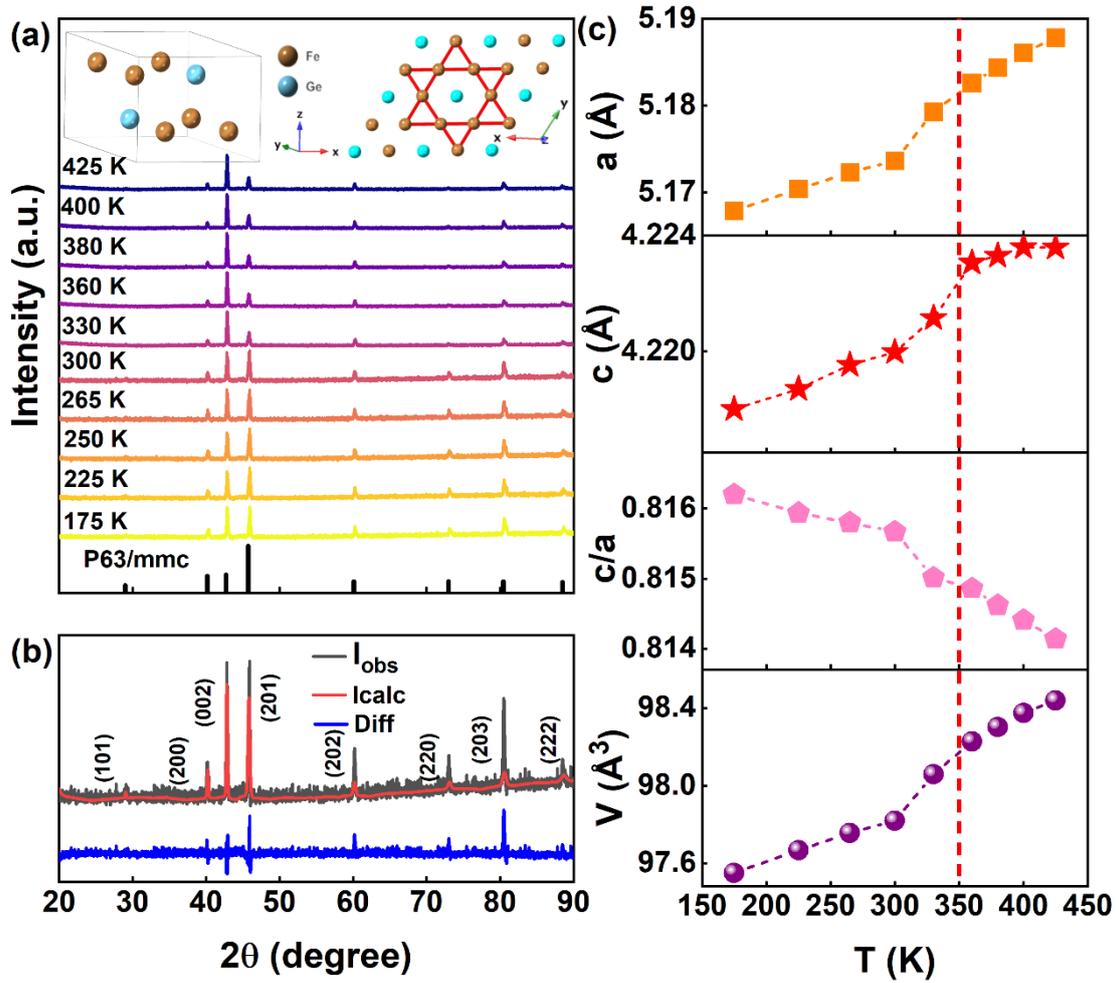

Fig. 1(a) The XRD patterns of Fe$_3$Ge ribbon samples at different temperatures, with the top inset showing the schematic diagram of the D0$_{19}$-type crystal structure of Fe$_3$Ge. (b) The refined XRD results of the room-temperature Fe$_3$Ge ribbon sample, where the black spectrum represents the experimental test results, the red spectrum represents the refined fitting results, and the blue spectrum represents the difference between the experimental and simulated values. (c) The lattice constants $a$ and $c$, the $c/a$ ratio, and the unit cell volume $V$ of the Fe$_3$Ge ribbon sample as functions of temperature.

Fig. 1(a) shows the XRD patterns of Fe$_3$Ge ribbon samples at different temperatures. Comparing with the black standard peaks of the D0$_{19}$-type Fe$_3$Ge and the XRD patterna at room temperature, no traces of other phases were found, indicating that the Fe$_3$Ge ribbon sample exhibits the D0$_{19}$ structure. The XRD patterns of Fe$_3$Ge ribbon sample at different temperatures suggest that the change in temperature did not induce any structural phase transition in the Fe$_3$Ge ribbon sample, which consistently maintained the D0$_{19}$ structure. The inset depicts the crystal structure of D0$_{19}$-type Fe$_3$Ge, where Fe atoms form a two-dimensional kagome lattice on the ab plane, while Ge atoms are located at the center of the kagome lattice, and two adjacent kagome Fe$_3$Ge layers stack along the c-axis. The XRD patterns at different temperatures were fitted using the Rietveld method, and the goodness of fit $\chi^2$ was less than 2 for all fitting results. Fig. 1(b) shows the refined XRD results of the Fe$_3$Ge ribbon sample at room temperature, where the black spectrum represents the experimental test results, the red spectrum represents the refined fitting results, and the blue spectrum represents the difference between the experimental and simulated values, it can be seen that the experimental XRD patterns are basically consistent with the fitting results. As shown in Fig. 1(c), we plotted the temperature dependence of lattice constants $a$, $c$, the $c/a$ ratio, and the unit cell volume $V$. We observed that both lattice constants $a$ and $c$ increase with temperature. However, lattice constant $a$ sharply increases after 350 K, while lattice constant $c$ gradually tends to saturate. According to literature reports[13,14], the spin reorientation temperature of $Fe_3Ge$ is 380 K. We predict that the anisotropic changes in lattice constants $a$ and $c$ after 350 K may be related to spin reorientation. The $c/a$ ratio decreases with increasing temperature, while the unit cell volume $V$ increases with temperature, with a change in slope occurring after 350 K. The increase in lattice constants $a$, $c$, and unit cell volume $V$ with temperature is attributed to thermal expansion of the lattice. In conclusion, we infer that $Fe_3Ge$ undergoes an isostructural phase transition.

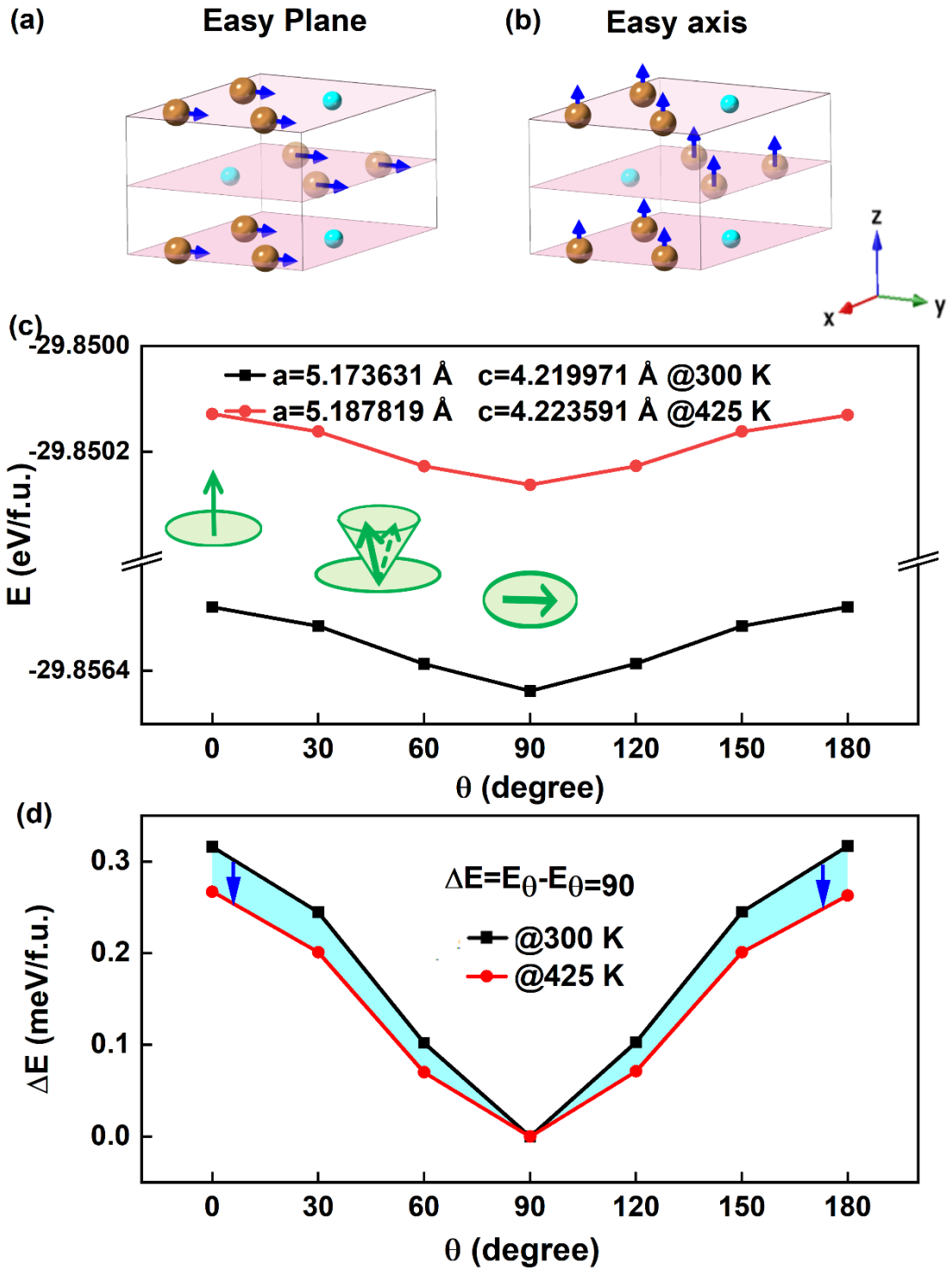

Fig. 2 Schematic diagram of the crystal structure and magnetic structure of Fe$_3$Ge with easy-plane (a) and easy-axis (b) magnetic anisotropy. In the diagram, brown spheres represent Fe atoms, light blue spheres represent Ge atoms; dark blue arrows indicate spin directions, and light red crystal planes represent the ab basal plane of the hexagonal structure. (c) The energy of Fe$_3$Ge under different magnetic configurations, where the angle $\theta$ is the angle between the magnetic moment of Fe$_3$Ge and the easy axis. The black square and red circle represent the energy calculation results based on the refined lattice constants obtained from XRD of Fe$_3$Ge ribbon samples at 300 K and

425 K respectively. The inset showing the schematic diagram the spin directions corresponding to different angles, where the light green plane represents the ab plane, the green arrows indicate the spin direction, and the light green cone represents the spin metastability. (d) The energy difference of $Fe_3Ge$ under different magnetic configurations and with spins in the easy plane as a function of $\theta$.

In order to further investigate the spin configurations of $Fe_3Ge$ at different temperatures, we used 350 K, where lattice constants exhibit anomalies, as a dividing line and performed two sets of first-principles calculations based on experimental lattice constants at 300 K and 425 K, which we named as 300 K and 425 K, respectively. Fig. 2(a-b) depict schematic diagrams of the crystal structure and magnetic structure of $Fe_3Ge$ with easy-plane and easy-axis magnetic anisotropy. In the diagrams, brown spheres represent Fe atoms, light blue spheres represent Ge atoms; dark blue arrows indicate spin directions, and light red crystal planes represent the ab basal plane of the hexagonal structure. Fig. 2(c) shows the energy of $Fe_3Ge$ under different magnetic configurations, where the angle $\theta$ is the angle between the magnetic moment of $Fe_3Ge$ and the easy axis. The black square and red circle represent the energy calculation results based on the refined lattice constants obtained from XRD of $Fe_3Ge$ ribbon samples at 300 K and 425 K respectively. The inset showing the schematic diagram the spin directions corresponding to different angles, where the light green plane represents the ab plane, the green arrows indicate the spin direction, and the light green cone represents the spin metastability. We found that in both sets of calculations, at $\theta = 90°$ corresponding to the easy plane, the energy is lowest, indicating the most stable magnetic configuration. However, the energy for all angles $\theta$ at 425 K is higher than at 300 K, suggesting a less stable magnetic configuration at 425 K. To further quantify the differences between 425 K and 300 K, as shown in Fig. 2(d), we calculated the energy corresponding to different $\theta$ and the energy difference of the $\theta = 90°$ easy plane. It can be observed that for all $\theta$ values, the energy difference at 425 K is lower than at 300 K, indicating a lower energy barrier between the easy plane and easy axis at the higher temperature above 350 K, making the magnetic moment of $Fe_3Ge$ less inclined to stabilize in the easy plane. Regarding the energy difference for the same $\theta$ and $\theta = 90°$ easy plane, the difference between 425 K and 300 K is only of the order of 0.01. However, in the study by Claudia Felser et al. on the $Mn_2YGa$ system where different main group elements replace Y, reported energy differences are at the order of 0.1[15]. This could be because 300 K is close to the lattice constants anomaly point at 350 K, guessing that at 300 K, the magnetic moment is not completely stable in the easy plane and is in a metastable state.

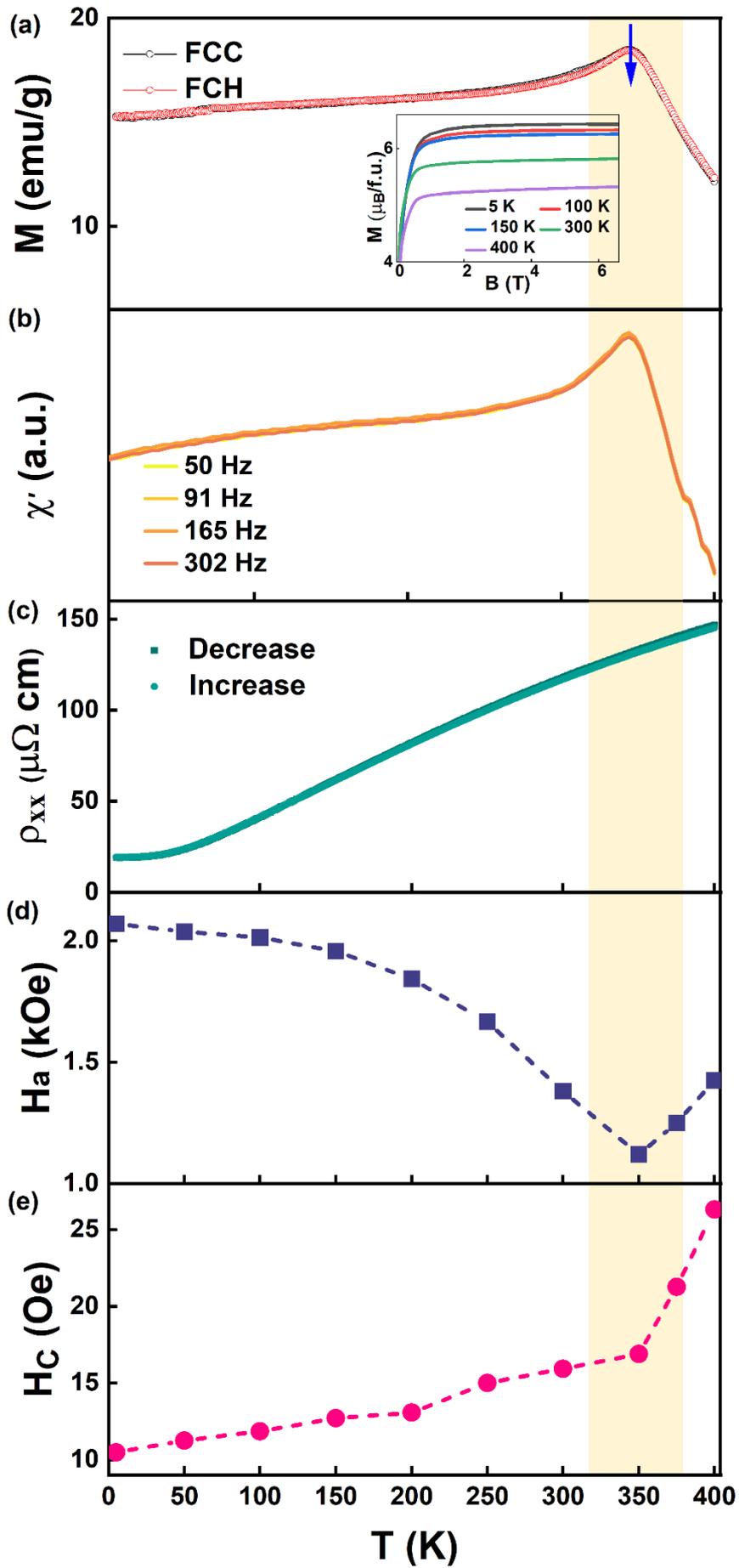

Fig. 3 (a) The temperature dependence of the magnetization intensity $M$ of field-cooled cooled (FCC) (light blue) and field-heated (FCH) (dark blue) samples under a 0.5 kOe magnetic field, with the transition temperature indicated by red arrows, inset showing the initial magnetization curves of the samples at different temperatures. (b) The temperature dependence of the real part of the AC magnetic susceptibility χ' of $Fe_3Ge$ samples at different frequencies (from 50 Hz to 302 Hz) with an AC amplitude of 10 Oe. (c) The temperature dependence of the longitudinal resistivity $\rho_{xx}$ at zero magnetic field. (d) The relationship between the magnetic anisotropy field $H_a$ and temperature. (e) The relationship between the coercive field $H_C$ and temperature.

We conducted macroscopic magnetic and electrical measurements to further investigate the magnetic configuration at different temperatures in the $Fe_3Ge$ ribbon samples. Fig. 3(a) shows the temperature dependence of the magnetization intensity of the $Fe_3Ge$ ribbon samples under field cooling (FC) and field heating (FH) in a 0.5 kOe magnetic field. "Peak-like" behavior was observed at 350 K in the FC and FH curves, consistent with the temperature at which lattice constants undergo anomalous changes. We suspect that the appearance of "peak-like" behavior may involve spin frustration or cluster glass[16], a long-range Néel temperature, or possibly spin reorientation[17]. As shown in Fig. 3(b), the temperature dependence of the AC magnetization also exhibits a peak at 350 K, and there is no shift with changing frequency, thus ruling out the possibility of spin frustration or cluster glass. If it were a long-range temperature, above the ordering temperature, it would be a paramagnetic phase. However, based on the initial magnetization curves at different temperatures in the inset of Fig. 3(a), the initial magnetization curves above the characteristic temperature of 350 K still exhibit saturated magnetization intensity, which does not align with the characteristics of a paramagnetic curve. According to literature reports, anomalies in longitudinal resistivity are typically observed near the characteristic temperature of long-range order[18]. Fig. 3(c) shows the temperature dependence of the longitudinal resistivity of $Fe_3Ge$ ribbon samples at zero magnetic field. It can be observed that the curve of longitudinal resistivity versus temperature exhibits a smooth transition across the entire temperature range, without any observed anomalies. Combined with the initial magnetization curves, the long-range order temperature is also ruled out. Therefore, it is concluded that 350 K is the spin reorientation temperature. However, Balagurov A M reported that the spin reorientation temperature of $Fe_3Ge$ is 380 K[14], which differs somewhat from the 350 K observed in our $Fe_3Ge$ ribbon samples. This difference may stem from the polycrystalline ribbon technique used in the sample preparation process.

The occurrence of spin reorientation is accompanied by changes in the magnetic anisotropy field and coercive field. Fig. 3(d) shows the relationship between the magnetic anisotropy field and temperature for $Fe_3Ge$ ribbon samples. As the temperature increases to 350 K, the magnetic anisotropy field gradually decreases. This may be due to the slow initial stage of the phase transition, which cannot fully counteract the isotropic decay effect caused by thermal fluctuations, resulting in a continuous decrease in the magnetic anisotropy field. However, after 350 K, the magnetic anisotropy field abnormally increases. This phenomenon is attributed to the fact that the magnetic moments tend to align more along the easy axis above 350 K, and the easy-axis anisotropy of the $D0_{19}$ structure is stronger than the easy-plane anisotropy, leading to the increase in the magnetic anisotropy field. Fig. 3(e) shows the relationship between the coercive field ($H_C$) and temperature for $Fe_3Ge$ ribbon samples.

Throughout the entire temperature range of the measurement, the coercive field remains at a few tens of Oersteds, indicating that the sample exhibits soft magnetic characteristics, consistent with reports in the literature[19]. The coercive field of Fe$_3$Ge increases by only about 5 Oe from 5 K to 350 K. However, when the temperature exceeds 350 K, the coercive field sharply increases. The variations of the magnetic anisotropy field and coercive field with temperature further confirm that 350 K is the spin reorientation temperature of the Fe$_3$Ge ribbon samples.

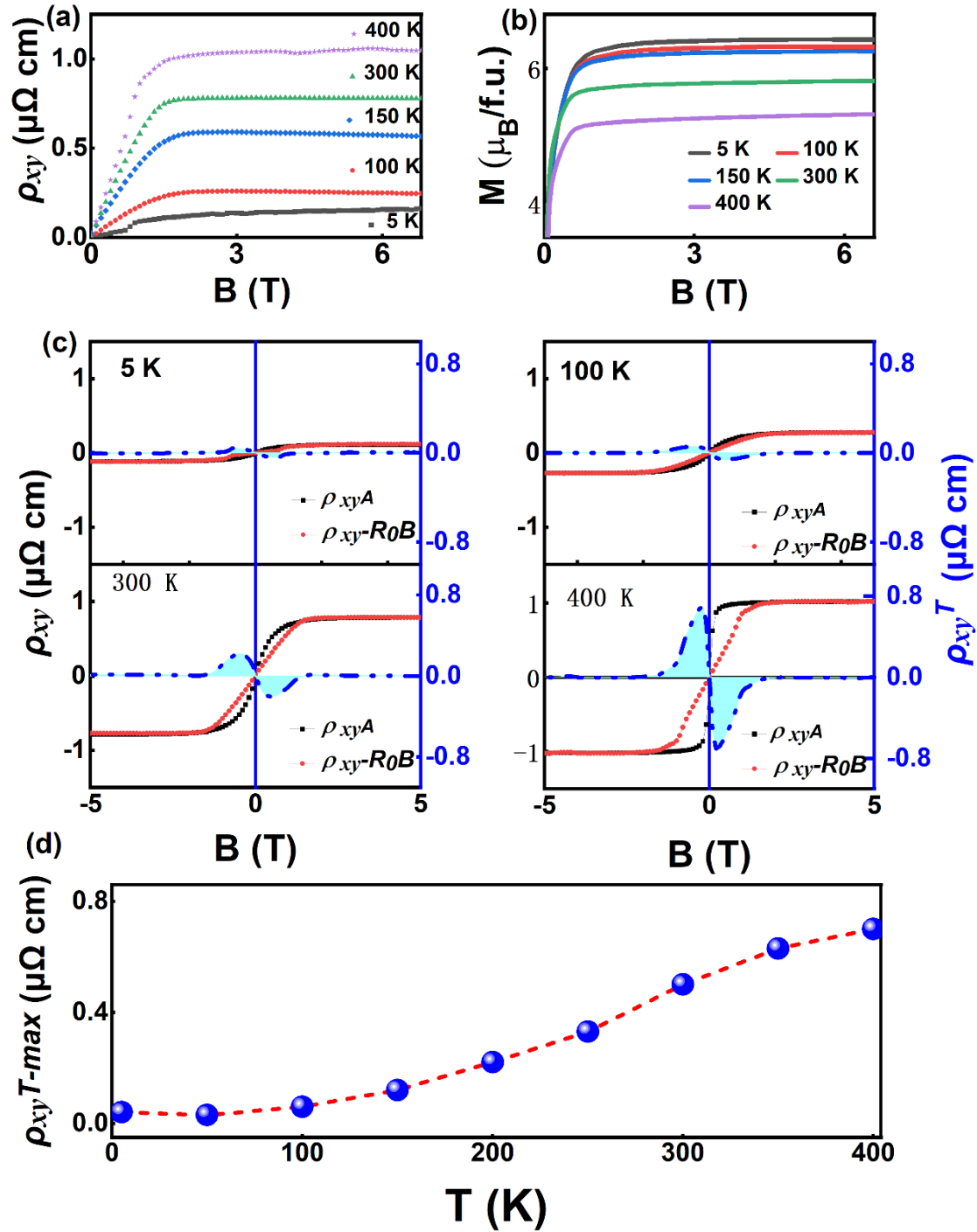

Fig. 4(a) The dependence of the total Hall resistivity $\rho_{xy}$ on magnetic field at different temperatures for Fe₃Ge ribbon samples. (b) Initial magnetization curves of Fe₃Ge ribbon samples at different temperatures. (c) Extraction process of the topological Hall resistivity, where the red curve represents the residual Hall resistivity after subtracting the normal Hall resistivity, the black curve represents the anomalous Hall resistivity, and the blue curve represents the topological Hall resistivity. (d) The temperature dependence of the maximum value of the topological Hall resistivity $\rho_{xy}^{T\text{-max}}$.

The occurrence of spin reorientation is highly likely to produce the topological Hall effect, so we conducted Hall resistivity tests. Fig. 4(a) shows the dependence of the total Hall resistivity $\rho_{xy}$ on magnetic field at different temperatures for Fe₃Ge ribbon samples. We found that within a wide temperature range from 5 K to 400 K, the value of $\rho_{xy}$ is non-zero, and the saturation value increases with temperature. Fig. 4(b) shows the initial magnetization curves of Fe₃Ge ribbon samples at different temperatures. All *M-H* curves exhibit typical ferromagnetic (FM) behavior and saturate at around 1 T magnetic field. Fig. 4(c) illustrates the extraction process of the topological Hall resistivity. Following the extraction method of topological Hall resistivity proposed by Li [20,21] and others, we utilized the formula $\rho_{xy} = \rho_{xy}^N + \rho_{xy}^A + \rho_{xy}^T = R_0 B + R_s \mu_0 M + \rho_{xy}^T$ to extract the topological Hall resistivity from the total Hall resistivity. Here, $\rho_{xy}^N$ represents normal Hall resistivity, $\rho_{xy}^A$ denotes anomalous Hall resistivity, $\rho_{xy}^T$ stands for topological Hall resistivity, $R_0$ and $R_s$ correspond to normal and anomalous Hall coefficients, respectively, *B* is the magnetic field, and *M* is the magnetization intensity. The red curve represents the residual Hall resistivity after subtracting the normal Hall resistivity, the black curve represents the anomalous Hall resistivity, and the blue curve indicates the topological Hall resistivity obtained by subtracting the anomalous Hall resistivity from the residual Hall resistivity. The topological Hall resistivity signal of our Fe₃Ge ribbon samples is similar to that of Fe₃Sn₂[11] and NdCo₅[22], exhibiting reversible loops under positive and negative fields. We observed that in the low magnetic field region, the topological Hall resistivity first shows a sharp increase, followed by a rapid decrease to zero. This is because in the high magnetic field region, the magnetic moments are aligned in a collinear ferromagnetic (FM) arrangement, resulting in the disappearance of topological Hall resistivity, leaving only normal and anomalous Hall resistivity. Below the spin reorientation temperature of 350 K, within the temperature range of 5 K to 300 K, the value of the topological Hall resistivity is relatively small. Above the spin reorientation temperature of 350 K, the value of the topological Hall resistivity significantly increases. To quantify the variation of the topological Hall resistivity at different temperatures, we extracted the maximum values of the topological Hall resistivity at different temperatures from Fig. 4(c) as shown in Fig. 4(d). With the increase in temperature, the topological Hall resistivity increases from 0.05 μΩ cm at 5 K to 0.69 μΩ cm at 400 K, approximately 14 times higher. Due to the easy-plane magnetic anisotropy of the Fe₃Ge ribbon sample at low temperatures, it transitions to easy-axis magnetic anisotropy as the temperature rises. In summary, we observed a significantly temperature-sensitive large topological Hall resistivity in our Fe₃Ge ribbon sample.

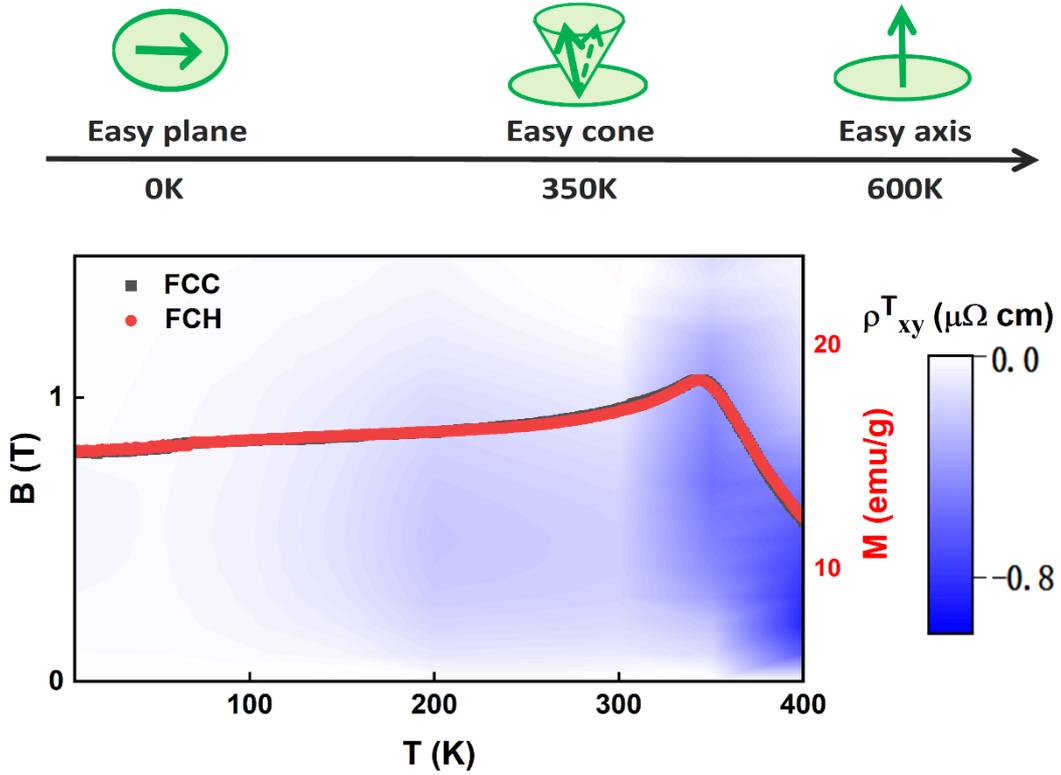

Fig. 5 (a) Schematic diagram of the magnetic structure of $Fe_3Ge$, quoted from Ref.[23]. (b) Phase diagram of the topological Hall resistivity dependence on magnetic field and temperature for the $Fe_3Ge$ ribbon sample. The black curve and red curve represent the temperature dependence of the sample magnetization intensity *M* under field-cooling (FCC) and field-heating (FCH) conditions at a 0.5 kOe magnetic field, respectively.

To further investigate the variation of the topological Hall resistivity with temperature in the $Fe_3Ge$ ribbon sample, as shown in Fig. 5(a), we combined first-principles calculations and magnetic testing analysis to draw schematic diagrams of the magnetic structure in different temperature ranges. The light green plane represents the ab plane, the green arrows indicate the spin direction, and the light green cone represents the spin metastability. As shown in Fig. 5(b), the phase diagram of the dependence of topological Hall resistivity on magnetic field and temperature is drawn. The colored rectangle on the right represents the range of topological Hall resistivity. The gradual transition from white to dark blue represents the absolute value of topological Hall resistivity gradually increases from 0. The black curve and red curve are the temperature dependence of the sample magnetization *M* under field cooling (FCC) and field heating (FCH) conditions under a magnetic field of 0.5 kOe, respectively. At low temperatures, the spins of the $Fe_3Ge$ ribbon sample tend to align along the easy-plane. As the temperature increases, the anisotropy of the easy-plane gradually weakens, reducing the energy barrier between the easy-plane and easy-axis. Therefore, under the influence of an external magnetic field, it becomes more favorable for spin non-coplanarity, leading to a gradual increase in the topological Hall resistivity. As the temperature further increases to 350 K, the peak temperature of both the FCC and FCH curves, spin reorientation occurs, gradually tilting the spins towards the c-axis. The topological Hall resistivity begins to increase rapidly, reaching its maximum value of 0.69 μΩ cm at 400 K.

**Conclusion**

In this work, we conducted a systematic study on the magnetic configuration and topological Hall resistivity of the $Fe_3Ge$ ribbon sample. Firstly, $Fe_3Ge$ polycrystalline ribbons were successfully prepared through arc melting and ribbon technology in an argon atmosphere. Variable temperature XRD spectra confirmed that there was no structural phase transition in the $Fe_3Ge$ ribbon sample at different temperatures, and it maintained the $D0_{19}$ structure throughout. By fitting the variable temperature XRD spectra, the temperature-dependent relationship of the lattice constants was obtained, revealing anisotropic changes in the lattice constants *a* and *c* after 350 K, indicating a change in the magnetic structure of the system. First-principles calculations indicate that at low temperatures, the spin tends to stabilize in the easy plane, while after 350 K, the spin favors the easy axis. A series of magnetic and electrical tests confirmed the occurrence of spin reorientation at 350 K. The occurrence of spin reorientation is highly likely to produce the topological Hall effect. The test results of the Hall resistivity reflect that the topological Hall resistivity increases with temperature after the spin reorientation temperature, reaching a maximum value of 0.69 μΩ cm at 400 K.

Combining first-principles calculations and magnetic testing analysis, it is found that in the $Fe_3Ge$ ribbon sample at low temperatures, the spins tend to align along the easy plane. As the temperature increases, the anisotropy of the easy plane gradually decreases, and the energy barrier between the easy plane and the easy axis diminishes. Therefore, under the influence of an external magnetic field, it is more favorable for the spins to be non-coplanar, leading to the emergence of the topological Hall resistivity. With further temperature increase, spin reorientation occurs with the spins tilting towards the c-axis. After the spin reorientation temperature, the topological Hall resistivity sharply increases, reaching a peak at 400 K. This work is the first to observe the topological Hall effect in $Fe_3Ge$ with spin reorientation, introducing a new member to the family of topological Hall effects.